\documentclass[12pt]{iopart} 

\newcommand\nice[1]{#1}    \newcommand\subm[1]{}   

\nice{\newcommand\dosingle[1]{#1}  \newcommand\dodouble[1]{ }} 
\subm{\newcommand\dosingle[1]{ }   \newcommand\dodouble[1]{#1}} 

\usepackage{graphicx}  

\usepackage{natbib}  
\usepackage{amsfonts}

 \usepackage{hyperref}
\nice{ 
\providecommand{\eprint}[1]{\href{http://arxiv.org/abs/#1}{{\tt [arXiv:#1]}}}

\providecommand{\url}[1]{\href{#1}{#1}}
}

\providecommand{\adsurl}[1]{} 

\def\SSS{Sect.~}

\providecommand\apj{ApJ}                 
\providecommand\apjs{ApJSupp}                 
\providecommand\aap{A\&A}            
\providecommand\mnras{MNRAS}
\providecommand\nat{Nature}
\providecommand\cqg{CQG}

\providecommand\prd{Phys.~Rev.~D}
\providecommand\physrep{Phys. Rep.}
\providecommand\BASI{Bull. Astr. Soc. India}

\providecommand\jams{J.~Amer.~Math.~Soc.}

\providecommand\Eqdot{{}}
\providecommand\Eqs{{}}
\providecommand\Figdot{Figure~}
\providecommand\Figs{Figures~}

\nice{ \newcommand\mycaptionfont{\protect\footnotesize} }

\def\gtapprox{\,\lower.6ex\hbox{$\buildrel >\over \sim$} \, }
\def\ltapprox{\,\lower.6ex\hbox{$\buildrel <\over \sim$} \, }
\def\propapprox{\,\lower.6ex\hbox{$\buildrel \propto\over \sim$} \, }

\def\arcs{\ifmmode {'' }\else $'' $\fi}     
\def\arcm{\ifmmode {' }\else $' $\fi}       

\def\fr7{7$ \hskip -0.9ex \vrule height0.8ex width0.8ex depth-0.73ex
                                                     \hskip0.1ex$}
\def\frtoday{Le\space\number\day\space\ifcase\month\or
  janvier\or f\'evrier\or mars\or avril\or mai\or juin\or
  juillet\or ao\^ut\or septembre\or octobre\or novembre\or 
d\'ecembre\fi\space \number\year}

\newcommand\hGpc{\mbox{$h^{-1}$ Gpc}}

\newcommand\rinj{r_{\mathrm{inj}}}  
\newcommand\rinjdimful{r_{\mathrm{inj}}'} 
\newcommand\Omk{\Omega_{\mathrm{k}}}
\newcommand\rhotot{\rho_{\mathrm{tot}}}
\newcommand\rhocrit{\rho_{\mathrm{crit}}}

\newcommand\rC{R_{\mathrm{C}}}

\newcommand\Fnaive{F_{\Gamma = 0}} 
\newcommand\FOm{F_{m,H}^{\Omega}} 
\newcommand\Frinj{F_{m,H}^{2\rinj}} 

\newcommand\tposttopo{t_{\mathrm{t}}}  

\newtheorem{hypothesis}{Hypothesis}

\newcommand\postrefereechanges[1]{#1} \newcommand\postrefereestart{ }  \newcommand\postrefereestop{ } 

\begin{document}

\title[A measure on compact FLRW models]{A measure on the set of compact
  Friedmann-Lema\^{\i}tre-Robertson-Walker models}

\author[B. F. Roukema \& V. Blanl{\oe}il]{Boudewijn F Roukema$^1$
and Vincent Blanl{\oe}il$^2$
\\
$^1$ 
Toru\'n Centre for Astronomy, Nicolaus Copernicus University,
ul. Gagarina 11, 87-100 Toru\'n, Poland  
\\
$^2$
D\'epartement de Math\'ematiques, Universit\'e de Strasbourg,
7 rue Ren\'e Descartes, 67084 Strasbourg cedex, France
}


\date{\frtoday}


\begin{abstract}
{Compact, flat Friedmann-Lema\^{\i}tre-Robertson-Walker (FLRW) models
  have recently regained interest as a good fit to the observed cosmic
  microwave background temperature fluctuations.  However, it is
  generally thought that a globally, 
  {exactly-flat} FLRW model is theoretically
  improbable.}
{Here, in order to obtain a probability space on the set $F$ of
  compact, comoving, 3-spatial sections of FLRW models, a physically
  motivated hypothesis is {proposed}, using the
  density parameter $\Omega$ as a {{\em derived}} rather than
  fundamental parameter.}
{We assume that the processes that select the 3-manifold 
  \protect\postrefereechanges{{also select}} a 
  global mass-energy and \protect\postrefereechanges{{a Hubble parameter}}.}
{The requirement that the local and global values of $\Omega$ are
  equal implies a range in $\Omega$ that consists of a single real
  value for any 3-manifold.  Thus, the obvious measure over $F$ is the
  discrete measure.}
{Hence, if the global mass-energy and Hubble parameter are
  \protect\postrefereechanges{a function of 3-manifold choice} among
  compact FLRW models, then probability spaces parametrised by
  $\Omega$ do not, in general, give a zero probability of a flat
  model.  Alternatively, parametrisation by a spatial size parameter,
  the injectivity radius $\rinj$, suggests the Lebesgue measure. In
  this case, the probability space over the injectivity radius implies
  that flat models occur almost surely (a.s.), in the sense of
  probability theory, and non-flat models a.s.\ do not occur.}
\end{abstract}


{\pacs{98.80.Jk, 04.20.Gz, 02.40.-k}}



\maketitle 



\newcommand\fomRcneg{
\begin{figure}  
\centering
\includegraphics[width=8cm]{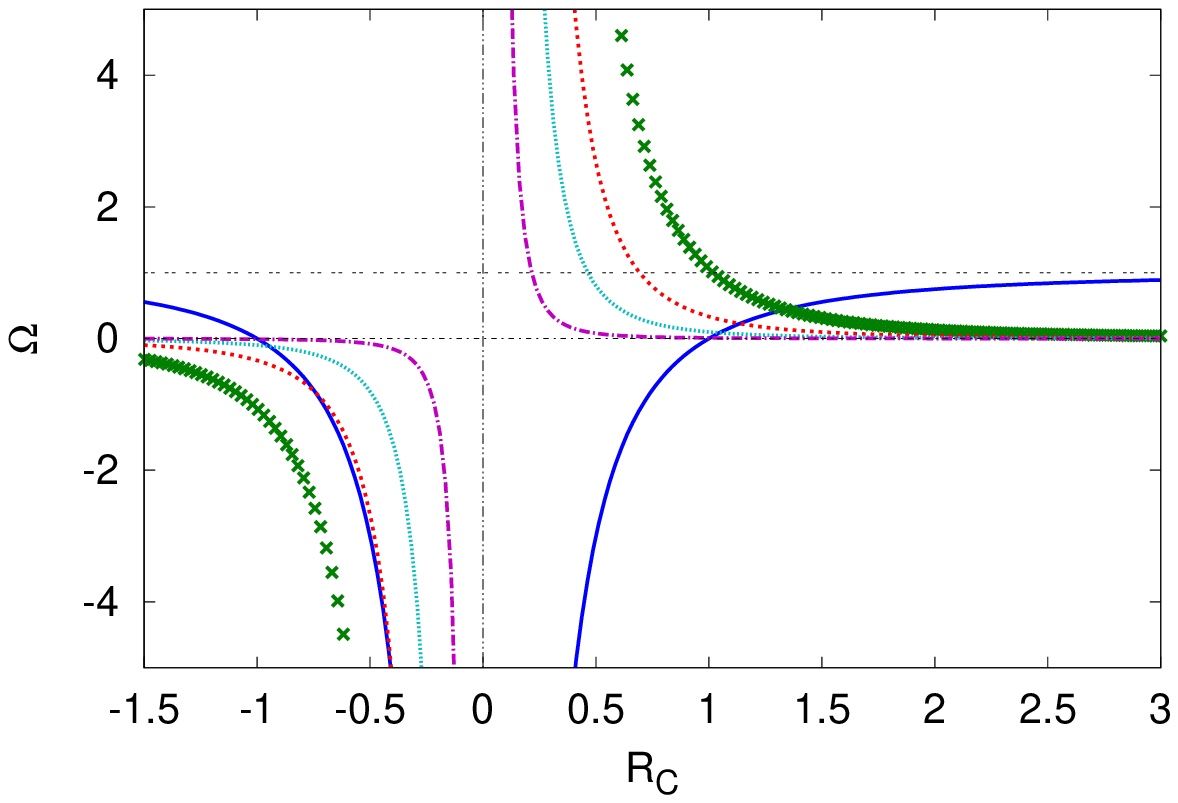}
\caption[comoving space]{ 
\mycaptionfont 
\protect\postrefereechanges{Example of local} 
and global dependence of the (total) density parameter $\Omega$
on the curvature parameter $\rC$ in hyperbolic, compact
FLRW models, given Hypothesis~\protect\ref{hyp-m-H-dependonM},
\protect\postrefereechanges{in the special case where $m(M)$ and $H(M)$
are constants independent of $M$.}
The local definition of the density parameter
 {\Eqdot}(\protect\ref{e-defn-omlocal}) gives the pair of curves that are symmetric
around $\rC=0$ and decrease without bound as $\rC \rightarrow 0$
(solid curves).
The global definition of the density parameter 
{\Eqdot}(\protect\ref{e-defn-omglobal}) is shown for 
dimensionless volume $V=0.943$ (``$\times$'' symbols; 
this is for the smallest-volume 
orientable hyperbolic 3-manifold, the Weeks-Matveev-Fomenko
manifold)
and for $V = 3.0, 10.0,$ and $100.0$.
The unphysical range
$\rC \le 0$ is shown for algebraic completeness, in order to show the cases
where one or more unphysical solutions with $\rC < 0$ occur.
The units are chosen so that 
\protect\postrefereechanges{$c=H(M)=8\pi \, G\, m(M) /3=1$, i.e. $\rC$ is in units
of $c/H(M)$}.
\label{f-omRcneg}
}
\end{figure}
} 

\newcommand\fomRcpos{
\begin{figure}  
\centering
\includegraphics[width=8cm]{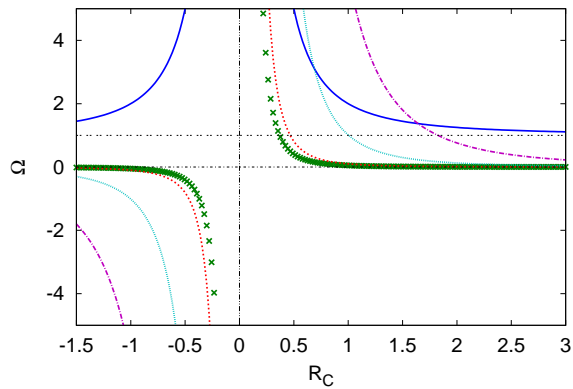}
\caption[comoving space]{ 
\mycaptionfont 
As for {\Figdot}\protect\ref{f-omRcneg}, for spherical models.
The local definition of the density parameter 
{\Eqdot}(\protect\ref{e-defn-omlocal}) 
gives the pair of curves symmetric
around $\rC=0$ that increase without bound as $\rC \rightarrow 0$ 
(solid curves).
The global definition of the density parameter 
{\Eqdot}(\protect\ref{e-defn-omglobal})
is shown for $V=2\pi^2$ (``$\times$'' symbols; 
this is for the largest spherical 3-manifold $S^3$)
and $V = \pi^2, \pi^2/10,$ and $\pi^2/60$.
The unphysical range $\rC \le 0$ yields no solutions.
\label{f-omRcpos}
}
\end{figure}
} 

\newcommand\fhex{
\begin{figure}  
\centering
\includegraphics[width=8cm]{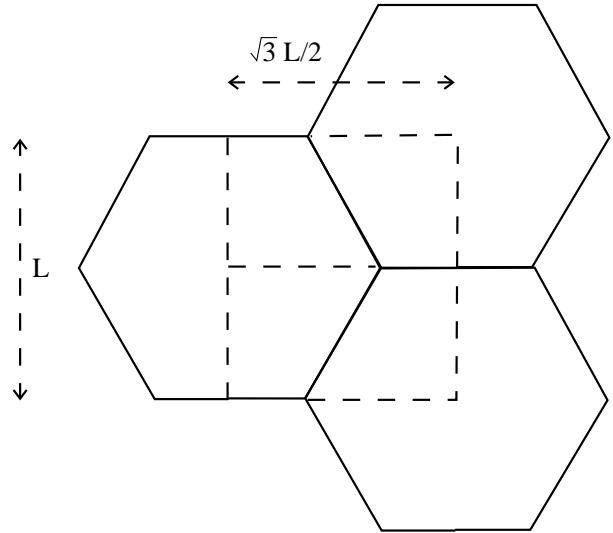}
\caption[]{ 
\mycaptionfont 
\protect\postrefereechanges{Orthogonal projection in $z$ direction of one
copy of the fundamental domain (FD) of a hexagonal-prism (third-turn or
sixth-turn) space illustrating its volume, given that the injectivity radius
(horizontal and vertical shortest closed geodesic) is $L$. 
The hexagonal area can be cut and paste into an $L \times \sqrt{3}L/2$ rectangle.}
\label{f-hex}
}
\end{figure}
} 

\newcommand\fHW{
\begin{figure}  
\centering
\includegraphics[width=8cm]{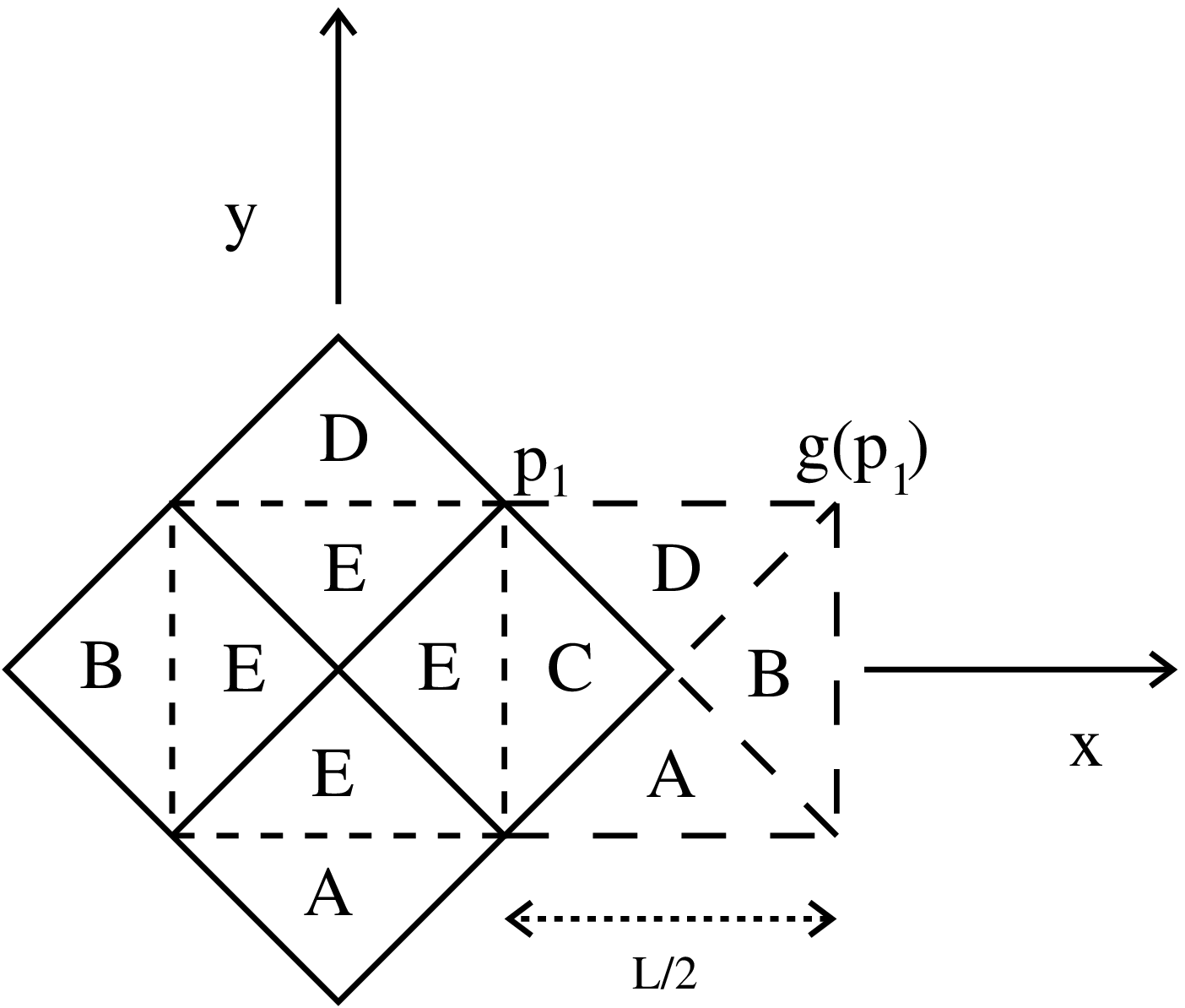}
\caption[]{ 
\mycaptionfont 
\protect\postrefereechanges{Orthogonal projection in $z$ direction of one
copy of the fundamental domain (FD) of the Hantzsche-Wendt space
illustrating its volume and injectivity radius. Looking
down on the FD from high $z$, four rhombic (diamond) faces,
AE, BE, DE, and CE,
are visible as projections to squares (solid lines).
The FD can be thought of as a cube of side length $L/2$ 
inscribed in the FD, surrounded by
lateral square-pyramidal extensions
A, B, C, and D, a superior square-pyramidal extension E, and a corresponding
one below the cube (not shown).  Positions to which some of these can be pasted
for volume calculation are shown by 
long-dashed lines. 
The $X$ and $Y$ axes' zero points are at the centre of 
the FD. Point $p_1$ is mapped to $g(p_1)$ by the holonomy $g$ 
(\protect\ref{e-defn-g}).}
\label{f-HW}
}
\end{figure}
} 


\section{Introduction}   \label{s-intro}

Recent empirical analyses regarding the shape of the comoving spatial
section of the Universe, assuming a perturbed
Friedman-Lema\^{\i}tre-Robertson-Walker (perturbed-FLRW) model, have
mostly focussed on the Wilkinson Microwave Anisotropy Probe (WMAP)
all-sky maps of the cosmic microwave background. The results are
presently inconclusive regarding both curvature and topology. The
curvature has been found to be ``nearly'' zero, i.e., comoving space
could be hyperbolic, flat or spherical. 
Writing the curvature
as an effective density parameter $\Omk := 1-\Omega$, where $\Omega$ is the total
density parameter,\footnote{The total density parameter is defined
$\Omega: = \rhotot/\rhocrit$, 
where $\rhotot$ 
and $\rhocrit$ are the
total density and the critical density, respectively. The latter
is $\rhocrit := 3H^2/(8 \pi G)$ where $H$ is the Hubble parameter and
$G$ the gravitational constant.}
none of the three possible
curvatures $k \in  \{0, \pm 1\}$
have been observationally excluded to high statistical
significance. It is true that the precision in upper limits
to $|\Omk|$ has improved by about one a half orders of magnitude
in the last decade and a half, and that the accuracy in 
$|\Omk|$ upper limits has similarly improved, but these limits do
not exclude any of $k \in  \{0, \pm 1\}$.


Results regarding topology, i.e. the fundamental group $\Gamma$ of 
holonomy transformations of the comoving spatial section,
are also inconclusive. Several groups have
found that the Poincar\'e dodecahedral space $S^3/I^*$,
with covering space $\widetilde{M} = S^3$ and fundamental group $\Gamma=I^*$,
is preferred over simply
connected infinite flat space
\citep{LumNat03,Aurich2005a,Aurich2005b,Gundermann2005,Caillerie07,RBSG08,RBG08}
in order to fit the WMAP data, while others exclude detectable cosmic
topology \citep{CSSK03,KeyCSS06}, or prefer simply connected infinite
flat space \citep{NJ07,LR08}.


{While} a well-established theoretical basis for cosmic
curvature exists (the Einstein field equations match the spatial
curvature of the Universe to its matter-energy density), there are
at present only hints as to what might constitute a theory of cosmic
topology.  Various quantum gravity aspects of cosmic topology include
the decay from pure quantum to mixed states \citep[e.g.,
][]{Hawking84a}, smooth topology evolution \citep[e.g., ][]{DowS98},
and work that could contribute to a physical motivation for deciding
which 3-manifold should be favoured by a theory of quantum cosmology
\postrefereechanges{\protect\citep{Masafumi96,CarlipSurya04}}.  A recent heuristic result is that
of the dynamical effect of cosmic topology in the presence of density
perturbations.  A residual weak-limit gravitational effect in the
presence of a density perturbation selects well-proportioned spaces
\citep{WeeksWellProp04} in general, and the Poincar\'e dodecahedral
space $S^3/I^*$ in particular, to be special in the sense of being
``better balanced'' \citep{RBBSJ06,RR09}. The agreement between the
residual gravity effect and many of the empirical analyses in choosing
the same space, the Poincar\'e space, suggests that this effect might
have been important in the early universe in selecting the comoving
3-manifold in which we live.

On the other hand, \citeauthor{Aurich07align}
\citep{Aurich07align,Aurich08a,Aurich08b,Aurich09a} have recently
carried out several studies showing that a {\em flat} compact model
(specifically, $T^3$) generally provides a better fit to the WMAP data
than infinite flat models.  This work provides one of the few ways of
potentially showing that the Universe is flat in the sense of being a
flat 3-manifold (apart from perturbations), if that is indeed the true shape
of the Universe, rather than a Poincar\'e space. Approaches that
ignore topology sometimes argue in favour of flatness while
simultaneously stating (correctly) that they cannot determine the curvature.  For
example,  \citet{Vardanyan09} calculate the odds ``in favour of a flat
Universe'', but also state that ``the [curvature] of the Universe is
not knowable'' if $\log_{10} |\Omk| < - 4$, and make no claim that
$\log_{10} |\Omk| \ge - 4$.
That is, they assert that either the
curvature has not yet been determined, or it is unknowable. In
contrast, use of the global properties of the comoving spatial section
potentially offers a way to determine the curvature, even if the
curvature is zero, as suggested by \citeauthor{Aurich09a}'s work,
and even in the (realistic) presence of perturbations.

{Although} flat, compact space models have been frequently
modelled, they have generally been disliked for fine-tuning
reasons, especially in relation to inflationary motivations and
scenarios. 
It is frequently stated that the probability of the spatial section of
the Universe, according to the exact-FLRW model, being exactly flat is
zero.  For example, in a discussion about common misconceptions
regarding ``standard universe models'' and inflation,
\citet{Ellis06Philo} states that being exactly flat ``requires {\em
  infinite} fine tuning of initial conditions; if say the two
millionth digit in the value of 
$\Omk$ is non-zero at any time, then the universe is not spatially
flat'' and that ``although the scale-free $k=0$ exponential case is
the model underlying the way many people approach the problem, it is
highly exceptional --- it is of zero measure within the space of all
inflationary FL models'' \citep[Sect.~2.8.1, 2.7.1,
  resp.,][]{Ellis06Philo}.  That is, if the measure $\mu$ on
the space $F$ of all exact-FLRW models is a probability measure,
i.e. if $\mu(F)=1$, then the probability of exact flatness is zero:
 $P(k=0) = \mu(k=0) = 0$.  Since inflationary scenarios only favour
near flatness, not exact flatness, the subclass of inflationary
scenarios $I \subset F$ can validly have $\mu(I) > 0 $, i.e. 
a positive probability, according to this 
{implicit definition of a measure
space $(F,\Sigma,\mu)$.}

However, is being exactly flat necessarily a physical possibility that
has zero measure?  \citet{Ellis99Carg} argue that finding a natural,
plausible, measure in the ``full space of cosmological space-times''
is an open problem. Previous suggestions of measures
{\citep[e.g.,][]{GibbHawk87,Coule95,EvrC95,KirchE03,GibbT08}} 
have typically only considered the
curvature of the spatial section and not the fact that it is a
3-manifold that has both a curvature and a topology.  In other words,
the set $F$ of exact-FLRW models is considered to be equivalent to
$\Fnaive := \{ \Omega \in \mathbb{R} : \Omega > 0 \}$, i.e. it is
parametrised only by the density parameter $\Omega$.  The measure is
implicitly assumed to be a simple function of the Lebesgue measure on
$\Fnaive$ that gives $\mu(\Omega = 1) = 0$ and a normalised measure
$P(\Fnaive) = 1$, so that $P(\Omega=1) = 0$.

\postrefereechanges{The fact that an FLRW model has both a curvature and a topology
is not a
new insight. Both curvature and topology have 
been mentioned quite clearly by several of the
founders of relativistic cosmology
\citep{deSitt17,Fried23,Fried24,Lemaitre31ell,Rob35}. 
The curvature and topology of FLRW models have usually been thought to be 
independent of one another, except 
that
the set of possible 3-manifolds
is divided between hyperbolic, flat, and spherical 3-manifolds.} 
Some work that does consider topology
includes that of \postrefereechanges{\protect\citet{Masafumi96}} and \citet{CarlipSurya04}.
\postrefereechanges{\protect\citet{Masafumi96}} defined a ``spectral distance'' function over pairs
of universes with (in general) different spatial sections, but did not
try extending this to a measure. \citet{CarlipSurya04} applied a
Hartle-Hawking ``no boundary'' path integral approach to the spatial
section that constitutes a boundary to a global space-time.
{Alternatively,
a Bayesian approach to measure spaces and the flatness problem, using Jaynes'
principle, finds that the flatness problem is not a problem \citep{EvrC95,KirchE03}.}

Here, a more elementary approach 
to obtaining a measure on the set of exact-FLRW models, considering
both the curvature and topology of the models,
is presented.  In
\SSS\ref{s-hypothesis}, a physically motivated hypothesis is proposed.
This hypothesis is used to determine the set of compact, 3-spatial
sections of exact-FLRW models where the density parameter
$\Omega$ is a {\em derived} rather than fundamental physical parameter.
Only compact models are considered, with the motivation that this is
physically more reasonable than infinite models.\footnote{For hyperbolic
models, replacing the condition ``compact'' by ``finite volume'', 
i.e. allowing finite volume models containing infinite spatial geodesics,
should not modify the results in this paper.}
The resulting set and the natural measure on it are presented in
\SSS\ref{s-omega-result}.  An alternative definition of the set, using
the injectivity {radius $\rinj$} ({half the length of the} 
shortest closed comoving spatial
geodesic), and the natural measure on it are discussed in
\SSS\ref{s-rinj-result}. Conclusions are given in \SSS\ref{s-conclu}.

For early references to cosmic topology see
\citet{Schw00}\footnote{English translation: \protect\citet{Schw98}.};
\citet{deSitt17,Fried23,Fried24,Lemaitre31ell} and \citet{Rob35}.  For
a short, modern introduction see \cite{Rouk00BASI}.  For reviews, see
\cite{LaLu95,Lum98,Stark98,LR99,BR99,RG04}.

\section{\protect\postrefereechanges{Parametrisation by $\Omega$}}

\subsection{\protect\postrefereechanges{Hypothesis}}

\label{s-hypothesis}

\postrefereestart

The implicit assumption that the space of
exact-FLRW models consists of $\Fnaive := \{ \Omega \in \mathbb{R} :
\Omega > 0 \}$ is a statement that $\Omega$ is a free parameter,
unconstrained by prior physics. One way of introducing some
physical motivation for a theory of a set of possible universes
is to extrapolate from known physics.

Local geometry is a fundamental and experimentally well-established
part of modern physics, via the Einstein field equations. It would be
physically reasonable that as a result of quantum gravity processes
or during (mostly) smooth early-universe topology evolution as
explored by \citet{DowS98}, some global physical properties of the Universe
in the FLRW approximation (not necessarily the density $\Omega$)
could be determined by a global geometrical property. 
The most obvious global geometrical property is {\em geometria situs}
\citep{Euler1736}, known today as topology. For exact-FLRW models,
the topological characteristic that is usually discussed is 
equivalent to the choice of constant curvature 3-manifold.

What {property or} properties could most reasonably be hypothesised
to be determined by the choice of topology? Although density (in an
exact-FLRW model) can be evaluated globally, a more fundamental
property in a classical, pre-relativistic sense, is mass.  In
pre-relativistic physics, conservation of mass in any closed system
is a fundamental principle. In FLRW cosmological models, let us
define the global (by volume), total (by component) non-relativistic
plus relativistic mass-energy in the 3-spatial section as $m :=
\rho_{\mathrm{tot}} V'$, where $\rho_{\mathrm{tot}}$ is the total
(by component) mass-energy density, and $V'$ is the volume of the
spatial section in physical units, if $k=0$, and as $m :=
\rho_{\mathrm{tot}} V \rC^3$, where $\rC$ is the curvature radius in
physical units and $V$ is the volume of the spatial section in units
of $\rC^3$, if $k\not=0$. Since, in general, $m$ is not conserved
with time, due to the change in frequency of (in particular) photons
with respect to locally comoving observers, this should be evaluated at
a given epoch.  

Another parameter that could reasonably depend on the choice of
3-manifold is the Hubble parameter $H$.  Thus, along with the above
definition of $m$, the following hypothesis is proposed.


\begin{hypothesis}
  The topological evolution processes in the early universe that
  determine the comoving {spatial} 3-manifold $M$ of an FLRW universe
  lead to, at a given {post-topology-evolution} epoch $\tposttopo$, a
  global, total\footnote{The term ``global'' refers to the whole of
    comoving space, while ``total'' refers to all of the matter-energy
    density components, such as baryons, non-baryonic dark matter,
    neutrinos, radiation, and dark energy.}  matter-energy $m(M) :=
  m(M)|_{\tposttopo}$, and a rate of expansion, i.e. the Hubble
  parameter $H(M) := H(M)\vert_{\tposttopo} = (\dot{a}/a)(M)
  \vert_{\tposttopo} $. 
  That is, $m$ and $H$ at the epoch $\tposttopo$ are functions 
  from the space of exact-FLRW models $F$ to $\mathbb{R}$, 
  written $m(M)$, $H(M)$ to denote this dependence.
  The total mass-energy $m(M)$
  is determined in the comoving frame.  The manifold $M$ is assumed to
  be compact and orientable, and to be fixed for all epochs at $t >
  \tposttopo$.
\label{hyp-m-H-dependonM}
\end{hypothesis}

\postrefereestop

{The epoch $\tposttopo$ is thought 
of here to be pre-inflationary if an inflationary scenario is
invoked, but neither requires nor rejects an inflationary epoch
{at}  $t > \tposttopo$.}
\postrefereechanges{To investigate possible measure spaces 
using Hypothesis~\ref{hyp-m-H-dependonM}, spatially curved and flat models
need to be considered separately.}

\subsection{Curved models}   \label{s-meth-curved}

For curved models, first let us consider the density parameter
$\Omega$ as a local parameter, i.e. as a limit towards a space-time
{point.}
In order to have a real, positive curvature radius $\rC$ for both negative
($k=-1$) and positive ($k=+1$) curvature, let us define $\rC \in \mathbb{R}$ 
{such that 
$\rC > 0$ and}
the Friedmann
equation is
\postrefereechanges{\begin{eqnarray}
 \Omega = 1 \pm \left( \frac{c}{H(M)\rC} \right)^2 & ,  & k = \pm 1 .
\label{e-defn-omlocal}
\end{eqnarray}}

In a compact, exact-FLRW model, $\Omega$ 
{also has a global meaning.}
Let us write the 
volume of a compact 3-manifold $M$ of constant, non-zero curvature
and (real) curvature radius $\rC > 0$ as
$V \rC^3$, i.e. $V$ is dimensionless. 
For example, 
the lens space $L(p,q)$ for $p,q \in \mathbb{Z}, 0 < q \le p/2,$ with
$p,q$ relatively prime \citep[e.g.,][]{GausSph01}, 
has volume $2\pi^2 \rC^3 /p$, so that 
$V[L(p,q)] =  2\pi^2 /p$.
In the hyperbolic case, the uniqueness of $V$ for a given fundamental group
$\Gamma$ was less obvious than for the spherical case, but follows {from}
\citet{Mostow68}'s rigidity theorem.\footnote{Extended to the non-compact,
finite volume case by Prasad.}
The definitions of the critical density and the density parameter give
\postrefereechanges{\begin{equation} 
    \Omega := \frac{\rho}{\rhocrit} = 
    \frac{m(M)}{V \rC^3} \; \frac{8\pi G}{3\, [H(M)]^2}.
    \label{e-defn-omglobal}
\end{equation}}

What values of $\Omega$ simultaneously satisfy 
{\Eqs}(\ref{e-defn-omlocal}) and (\ref{e-defn-omglobal})?
Clearly, the local and global values of the density parameter
must be equal in an exact-FLRW model. Hence, 
\postrefereechanges{\begin{equation}
    1 \pm \left( \frac{c}{H(M)\rC} \right)^2 
    = \frac{m(M)}{V \rC^3} \; \frac{8\pi G}{3\, [H(M)]^2},
    \label{e-local-global-omega}
\end{equation}
where} $\pm$ correspond to curvatures $k = \pm 1$, respectively, as above.
The range of $\rC$ values that satisfy this equation, 
and in turn, 
the values of $\Omega$ that satisfy both 
{\Eqs}(\ref{e-defn-omlocal}) and (\ref{e-defn-omglobal}) for a given pair 
\postrefereechanges{$[m(M),H(M)]$,}
are presented in \SSS\ref{s-omega-result}.


\fomRcneg

\fomRcpos


\subsection{Zero spatial curvature}  \label{s-res-zero}

The equivalent of {\Eqdot}(\ref{e-defn-omlocal}) for a flat, compact, exact-FLRW model
is 
\begin{equation}
\Omega=1.
\label{e-defn-omlocal-flat}
\end{equation}
Hence, the equivalent of {\Eqdot}(\ref{e-defn-omglobal}) is not needed in order
to determine the valid range of $\Omega$ for flat models.


\subsection{The set $\FOm$ of FLRW models parametrised by $\Omega$}
\label{s-omega-result}

\postrefereechanges{The set $F$ of compact, exact-FLRW models is considered here
to be 
the set of compact, comoving spatial sections that have an FLRW metric.
A superscript $\Omega$ is used to indicate the parameter over which a measure space is to 
be defined. The subscripts $m,H$ indicate that Hypothesis~\ref{hyp-m-H-dependonM} 
is assumed.}

Equation~(\ref{e-local-global-omega}) is a cubic equation in $\rC$.
The discriminant is
\postrefereechanges{\begin{equation}
\Delta = -324 V^2 H(M)^2 ( \pm c^6 V^2 + 48 \pi^2 G^2 m(M)^2 H(M)^2).
\label{e-discriminant}
\end{equation}}
For positive curvature, $\Delta < 0$, so there is only one real root for $\rC$.
For negative curvature, 
{there can be up to three distinct, real
roots, but from {\Eqs}(\ref{e-defn-omlocal}) and (\ref{e-defn-omglobal}), it is
clear that only one of these is positive.}
This is illustrated in {\Figs}\ref{f-omRcneg} and
\ref{f-omRcpos}.  The multiple root or the two additional roots that
can occur in the hyperbolic case ({\Figdot}\ref{f-omRcneg}) occur for $\rC
<0$, which is unphysical given the definition of $\rC$ used
here.\footnote{An alternative definition could use imaginary values of
  $\rC$ for the hyperbolic case.}

\postrefereechanges{For illustrative purposes, we can consider
the case where $m(M)$ and $H(M)$ are equal for all 3-manifolds.}
Figure~\ref{f-omRcneg} shows that \postrefereechanges{in this case,} 
for hyperbolic
3-manifolds, 
\postrefereechanges{there is} a maximum possible curvature radius $\rC$, i.e. 
a maximum possible density parameter $\Omega$,
{at $\tposttopo$.}
These maxima 
are attained for the smallest-volume hyperbolic 3-manifold. If we consider only
orientable, hyperbolic 3-manifolds, then the Weeks-Matveev-Fomenko space
[m$003(-3, 1)$
in the {\sc SnapPea} census\footnote{The file
\url{/usr/share/snappea/ClosedCensusData/}
\url{ClosedCensusInvariants.txt} of 
version 3.0d3-20 of {\sc SnapPea} is referred to in this paper.}]
of volume $V\approx 0.943 $ is the smallest-volume orientable hyperbolic 3-manifold 
\citep{WeeksPhD85,MatvFom88,Gabai07}, shown here by ``$\times$'' symbols.
Larger volume hyperbolic 3-manifolds give the continuous curves in the figure
[shown {for $V \approx 3.0$, e.g. m$117(-5,2)$ in the {\sc SnapPea} census,}
{and for $V=10.0, 100.0$}], which successively give lower $\rC$ and $\Omega$.
In {\Figs}\ref{f-omRcneg} and \ref{f-omRcpos}, 
\postrefereechanges{not only are $m(M)$ and $H(M)$ fixed, but the
displayed ranges of numerical values of $\rC$ and $\Omega$ 
have been chosen for convenience only,
i.e. with the units $c=H(M)=8\pi\, G\, m(M) /3=1$.
Elsewhere, these parameters are considered to retain
their physical units.}

\postrefereechanges{Similarly,}
Figure~\ref{f-omRcpos} shows that for the spherical case, 
\postrefereechanges{fixed values of $m(M)$ and $H(M)$ independent of $M$}
imply a minimum
possible curvature radius $\rC$ as the choice of 3-manifold
(i.e. $\Gamma$) is varied, which corresponds to a maximum possible density
parameter $\Omega$.  These are attained for the largest-volume
spherical 3-manifold, i.e. $S^3$ itself, of volume $V = 2\pi^2$.
While both signs of the curvature give upper limits to $\Omega$ for 
\postrefereechanges{fixed $m(M)$ and $H(M)$} 
as $\Gamma$ varies, the hyperbolic case prevents $\Omega$ from
approaching the flat case, while the spherical case allows $\Omega$ to
approach arbitrarily close to flatness as the volume decreases.
The volume $V=\pi^2/60$ is that of the Poincar\'e dodecahedral space.

\postrefereechanges{More generally, i.e. without fixing $m(M)$ and $H(M)$,}
the unique physical solution for $\rC$ in both cases is
\begin{eqnarray}
{\rC}_{\pm}(M) 
&=&  
 {{\left(
      \sqrt{ \pm c^6{} V^2+48{} \pi^2{} m^2{} G^2{} H^2}  
      +4{} \sqrt{3}{} \pi {} m{} G{} H
      \right)^{{{2}\over{3}}}
      \mp c^2{} V^{{{2}\over{3}}}}
    \over
        {\sqrt{3}{} H{} V^{{{1}\over{3}}} 
          {} \left(  
          \sqrt{ \pm c^6{} V^2+48{} \pi^2{} m^2{} G^2{} H^2}
          +4{} \sqrt{3}{} \pi{} m{} G{} H
          \right) ^{{{1}\over{3}}}  
  }} , 
\label{e-RC-solution}
\end{eqnarray}
where the dependence of $\rC$ on the manifold $M$ occurs through the 
\postrefereechanges{dependences
$V(M), m(M),$ and $H(M)$}.

The density parameter for a curved FLRW model can be written 
using either {\Eqdot}(\ref{e-defn-omlocal}) or (\ref{e-defn-omglobal}). 
The former gives
\postrefereechanges{\begin{equation}
\Omega_{\pm}(M) = 1 \pm \left[ \frac{c}{H(M) {\rC}_{\pm}(M)} \right]^2,
\label{e-omega-curved}
\end{equation}
where} $M = \mathbb{H}^3/\Gamma$ or $M = S^3/\Gamma$,
and ${\rC}_{\pm}$ is given in {\Eqdot}(\ref{e-RC-solution}).
As noted above {\Eqdot}(\ref{e-defn-omlocal-flat}), the solution in the
flat case is 
\begin{equation}
\Omega_{k=0}(M) = 1.
\label{e-omega-flat}
\end{equation}

Together, these give the following result. For any given manifold $M$,
Hypothesis~\ref{hyp-m-H-dependonM} implies a single value of $\Omega$ rather {than}
a continuous interval of possible values of $\Omega$. In other words, 
{\em there is no freedom in choosing $\Omega$ once $M$ has been selected.}
Since the set of constant curvature manifolds $\{ M \}$ is countable
\citep[e.g.,][]{LaLu95},
we can write $\FOm$ as the countable (infinite) set
\begin{equation}
\FOm = \{ \Omega(M) : M \in F \},
\label{e-FOm-set}
\end{equation}
where $\Omega(M)$ is given by {\Eqs}(\ref{e-omega-curved}) and (\ref{e-omega-flat}).

\subsection{Measure spaces and probability spaces}
\label{s-omega-measures}


Let us choose the obvious $\sigma$-algebra, $2^{\FOm}$, i.e. 
the set of all subsets of $\FOm$, and any discrete measure $\mu$ on $\mathbb{R}$
satisfying
\begin{eqnarray}
\mu(\Omega) > 0  ,& \; & \Omega \in \FOm \nonumber \\
\mu(\Omega) = 0  ,&\; & \Omega \not\in \FOm .
\label{e-FOm-possible-measures}
\end{eqnarray}
Then $(\FOm, 2^{\FOm},\mu)$ is a measure space over the 
density parameter $\Omega$ of exact-FLRW
models of the Universe. 
It is clear that there is no physical requirement that $\mu(\Omega=1)$ be zero,
i.e. there is no requirement that the measure of the case of a flat FLRW model 
be zero.

\postrefereechanges{For a probability space $(\FOm, 2^{\FOm},\widehat{\mu})$
to be made starting from the measure $\mu$,}
it would be required that  
$\widehat{\mu}(\FOm) = 1$. 
Let the manifolds be enumerated $\{M_i\}_{i=1}^{\infty}$ and
$\Omega(M)$ as given by {\Eqs}(\ref{e-omega-curved}) and (\ref{e-omega-flat})
be enumerated $\Omega_j$, where 
a non-bijective function $j(i)$ is provided so that
$j_1 \not= j_2 \Rightarrow \Omega_{j_1} \not= \Omega_{j_2}$
{i.e.}
manifolds of equal $\Omega$ are included as a single case with respect
to the measure. (Otherwise, $\widehat{\mu}(\Omega_j)$ would be multiply defined
for some values of $\Omega_j$, e.g. $\Omega_j=1$.)
Any convergent series 
$\{x_j\}_{j=1}^{\infty}$ 
with $\Sigma_j x_j =1$ 
now satisfies
\begin{eqnarray}
\widehat{\mu}[\Omega_j(M_i)] = x_j  ,&\; & M_i \in F \nonumber \\
\widehat{\mu}(\Omega) = 0  ,&\; & \Omega \not\in \FOm ,
\label{e-FOm-possible-prob}
\end{eqnarray}
{giving} a probability space $(\FOm, 2^{\FOm},\widehat{\mu})$.  

For example, this would be satisfied 
by 
\begin{eqnarray}
  j=1, \; x_j = 1/\mathrm{e} 
  ,&&  1 \le i \le 6 \\
  j\ge 2, \; x_j = 1/[ (j-1)!\; \mathrm{e}] 
  ,&&  i \ge 7 \nonumber \\
  M_i = \mathrm{E}_{i}  
  ,&& 1 \le i \le 6,
\label{e-example-enumeration}
\end{eqnarray}  
where $\mathrm{E}_i$ are the six compact, \postrefereechanges{flat}, 
orientable 3-manifolds as
labelled in 
\postrefereechanges{Table~I of \citet{RiazFlat03}}, and the 
non-flat, compact, orientable 3-manifolds are enumerated by $i \ge 7, j(i) \ge 2$.
This gives
\begin{equation}
  \widehat{\mu}(\Omega = 1)  = 
  \widehat{\mu}(\Omega_1)  = 
  x_1 =
  1/\mathrm{e} \approx 
  0.37 > 
  0
\label{e-FOm-example}
\end{equation}
{That is, given 
this example of $\{x_j\}$,
the probability of a flat space would be} about 37\%,
{i.e. strictly greater than zero.}
No physical motivation is suggested for this particular
choice of $\{x_j\}$ and enumeration $\{M_i\}$ of the 3-manifolds. 
This choice of $\{x_i\}$ and corresponding (partial) enumeration
 $\{M_i\}$ are presented only as a numerical example of how 
$P(\Omega=1)  > 0$ is possible, with
$P(\Omega=1) \equiv \widehat{\mu}(\Omega=1) = 1/\mathrm{e}$.
{It would remain possible
to choose $\{x_j\}$ in a way that gives 
$P(\Omega=1) =0$,
but it would be not be necessary to do so.}
Additional physical assumptions would be required to determine what
{the normalised discrete measure $\widehat{\mu}$ should be. 
Alternatively, $\Omega$ might not be a physically useful choice
for defining a measure space. 
The construction in \SSS\ref{s-rinj-result} suggests one alternative.}



\subsection{How can $\Omega=1$ {{\em not}} be a case of measure zero?}
\label{s-disc-why-flat-not-measure-zero}

Fine-tuning arguments have played a role in cosmology for (at least)
several decades. It may seem strongly counterintuitive that exactly
$\Omega=1$ models do not necessarily constitute a case of measure
zero, since the number 1 is a single real number on the continuous
real number line.  How does 
{the topology of spatial sections} invalidate this argument?
The reason is that while topology in some sense might appear to give
``additional'' parameter freedom to the FLRW models, 
curvature can be thought of as a type of rigidity that together with
topology reduces a continuous interval in the possible range of $\Omega$ 
to a set of discrete points.

\postrefereechanges{Thus}, the hypothesis presented here leads to a physical
link between curvature and topology. This link
is represented algebraically
in {\Eqdot}(\ref{e-local-global-omega}). This equation
represents a {\em physical relation between curvature and topology implied
by Hypothesis~\ref{hyp-m-H-dependonM}.} For a given compact manifold $M$, the
relativistic
mass-energy 
\postrefereechanges{$m(M)$} must be distributed uniformly throughout the space,
expanding at the given rate 
\postrefereechanges{$H(M) = (\dot{a}/a)(M)$}, in a way that
simultaneously gives the curvature implied {\em locally} at each 
\postrefereechanges{point in comoving space}
by the Friedmann equation and 
fills the volume implied by the {\em global}
shape of the manifold. The local and global roles of the density $\Omega$ 
do not have the freedom to be unrelated to one another if 
\postrefereechanges{$m$ and $H$ are determined by the choice of 3-manifold}.

The full mathematics that underlies this physical argument has only become
known in the second half of the twentieth century, and some details even
more recently. The complete classification of spherical 3-manifolds
and \citet{Mostow68}'s rigidity theorem for compact, hyperbolic 3-manifolds play
a role by implying a unique volume $V$ for any given curved,
compact 3-manifold. A proof that the Weeks-Matveev-Fomenko manifold
is the smallest-volume orientable hyperbolic 3-manifold was only published 
in {2009} \citep{Gabai07}. 

It is this physical relation between $\Omega$ being both local and
global, which is a result of the rigidity of (constant curvature)
curved 3-manifolds, that reduces the apparent continuum of values of
$\Omega$ in the curved case to a discrete set.
The set of values of $\Omega$ in the flat case is 
already discrete, since it contains just one member, $\Omega=1$.
Hence, the full set of allowed values of $\Omega$ is discrete, and
the natural measure is a discrete measure $\mu$, as 
indicated in {\Eqdot}(\ref{e-FOm-possible-measures}),
 rather than the Lebesgue measure.  This is why 
$\Omega =1$ need not be a case of measure zero.

\section{\protect\postrefereechanges{Parametrisation by a one-dimensional size parameter}} 
\label{s-rinj-result}

Moreover, while Hypothesis~\ref{hyp-m-H-dependonM} reduces the
freedom of $\Omega$ in curved FLRW models, it implies {\em more} parameter
freedom in flat FLRW models 
for other parameters.

In the 3-torus model $T^3$, there are (at least) three parameters
required to define the fundamental domain.  Following \citet{LLL96},
we can write the size of the domain as $L_{\mathrm{a}} L_{\mathrm{e}} L_{\mathrm{u}}$.  For orthogonal
side-lengths of the fundamental 
{parallelepiped} of a $T^3$ model, the
definition of the critical density gives
\postrefereechanges{\begin{equation}
L_{\mathrm{a}} L_{\mathrm{e}} L_{\mathrm{u}} = \frac{8 \pi\,G\, m(T^3) }{3 [H(T^3)]^2},
\label{e-defn-omglobal-T3}
\end{equation}
as} for {\Eqdot}(\ref{e-defn-omglobal}).  This equation was ignored in
\SSS\ref{s-res-zero}, since it provides no constraint on $\Omega$. The
Friedmann equation in this case is {\Eqdot}(\ref{e-defn-omlocal-flat}),
which provides no constraint on the global parameters $L_{\mathrm{a}}, L_{\mathrm{e}}, L_{\mathrm{u}}$.
Hence, two of these parameters are \postrefereechanges{free.}
In general, the angles between the faces of
the fundamental domain constitute additional free geometrical
parameters, but for $L_{\mathrm{a}}, L_{\mathrm{e}}, L_{\mathrm{u}}$ defined orthogonally, 
these angles do not affect the volume of the manifold.

The parameter freedom for compact, flat FLRW models has frequently
been thought of as an empirically undesirable property, since it makes
the models easier to fit to observations, reducing the models'
falsifiability.  This freedom is not total.  In the presence of an
inhomogeneity, the residual gravity effect \citep{RBBSJ06,RR09}, can
be invoked as a motivation for $L_{\mathrm{a}} \approx L_{\mathrm{e}} \approx L_{\mathrm{u}}$.

\subsection{The set $\Frinj$ of FLRW models parametrised by $2\rinj$}

However, without invoking the residual gravity effect, the
parametrisation of compact, flat FLRW models can be made in a way that
also applies to compact, curved FLRW models, by using a size
parameter. Here, we propose twice the injectivity 
radius, i.e., the length of the
shortest closed comoving spatial {geodesic,} 
{written $2\rinj \rC$, 
where $2\rinj$ is dimensionless, in  
the curved case, 
and $2\rinjdimful$,} including a length
dimension, in the flat case.
This underestimates the
full parameter freedom of the flat case, providing a
conservative approach to constructing a measure space.

Equation~(\ref{e-defn-omglobal-T3}) implies an upper limit to 
$2\rinjdimful$.
This occurs for the regular $\mathrm{T}^3$ model,
in which $L_{\mathrm{a}} =L_{\mathrm{e}}=L_{\mathrm{u}}$, so that  
\postrefereechanges{\begin{equation}
 2\rinjdimful = \min(L_{\mathrm{a}},L_{\mathrm{e}},L_{\mathrm{u}}) = 
\left\{ \frac{8 \pi\,G\, m(\mathrm{T}^3) }{3 [H(\mathrm{T}^3)]^2} \right\}^{1/3}.
\end{equation}}
\postrefereechanges{Increasing} any of the dimensions decreases at least one of the other dimensions,
forcing $2\rinjdimful$ to decrease.
A reasonable empirical lower bound for 
$2\rinjdimful$ can also be set, 
e.g. a comoving scale of $\sim 10${\hGpc}.
\postrefereechanges{\citet{Aurich08a} estimate $2\rinj = 11.5\pm 0.3${\hGpc} from 
applying the cross-correlation Monte Carlo Markov chain method
\citep{RBSG08,RBG08} to the 5-year WMAP data for a $\mathrm{T}^3$ model.}
Hence, for the $\mathrm{T}^3$ and related compact models with 
rectangular-prism fundamental domains
\postrefereechanges{\begin{eqnarray}
  10 a(\tposttopo){\hGpc} 
  & \ltapprox &   2\rinjdimful
  \nonumber \\
  &=& \min(L_{\mathrm{a}},L_{\mathrm{e}},L_{\mathrm{u}})  
  \nonumber \\
  & \le &
  \left( \frac{8 \pi\,G\, }{3} \right)^{1/3}
  \max_{i=1,2,3}\left\{ \frac{m(\mathrm{E}_i) }{[H(\mathrm{E}_i)]^2} \right\}^{1/3},
\label{e-rinj-T3}
\end{eqnarray}
where E$_1$, E$_2$, E$_3$ follow Table~I of \citet{RiazFlat03}.
The compact flat models with hexagonal-prism fundamental domains, E$_4$ and
E$_5$,
have the corresponding limit
\begin{eqnarray}
  10 a(\tposttopo){\hGpc} 
  & \ltapprox &   2\rinjdimful
  \nonumber \\
  & \le &
    2^{1/3} 3^{-1/6} \left( \frac{8 \pi\,G\, }{3} \right)^{1/3}
  \max_{i=4,5}\left\{ \frac{m(\mathrm{E}_i) }{[H(\mathrm{E}_i)]^2} \right\}^{1/3}
\label{e-rinj-hexprism}
\end{eqnarray}
(\ref{s-app-hex}).
The other compact, orientable, flat model, i.e. the 
Hantzsche-Wendt space, E$_6$, has a smaller injectivity radius, 
maximised in the regular case, giving
\begin{eqnarray}
  10 a(\tposttopo){\hGpc} 
  & \ltapprox &   2\rinjdimful
  \nonumber \\
  & \le &
    2^{-1/3} \left( \frac{8 \pi\,G\, }{3} \right)^{1/3}
  \left\{ \frac{m(\mathrm{E}_6) }{[H(\mathrm{E}_6)]^2} \right\}^{1/3}
\label{e-rinj-HW}
\end{eqnarray}
(\ref{s-app-HW}).}

Hence, over all curvatures, we can classify the 
full set of compact
FLRW models by $2\rinjdimful$ for the flat case and 
$2\rinj \rC$ for the curved cases.
$\Frinj$ can now be written
\begin{eqnarray}
  \Frinj 
  &=& 
  \bigg\{ 2\rinjdimful : 
  10 a(\tposttopo)  {\hGpc}
  \; < 2\rinjdimful 
  \le 
  \;
\postrefereechanges{
 \left( \frac{8 \pi\,G\, }{3} \right)^{1/3} \;\eta
}
  \bigg\}  \nonumber \\
  && \bigcup \;
  \left\{ 2\rinj(M) \, 
  {\rC(M)} : M \in F, 
  \widetilde{M} \in \{ \mathbb{H}^3 , S^3 \}
  \right\},
  \label{e-Frinj-set}
\end{eqnarray}
\postrefereechanges{where
  \begin{eqnarray}
    \eta &:=& \max \left(
    \left\{ 
    \left\{ 
    \frac{m(\mathrm{E}_i) }{[H(\mathrm{E}_i)]^2} \right\}^{1/3}, i=1, 2, 3
    \right\}  \right.
    \nonumber \\
    && \bigcup \;
    \left\{ 
     2^{1/3} 3^{-1/6} 
    \left\{ 
     \frac{m(\mathrm{E}_i) }{[H(\mathrm{E}_i)]^2} \right\}^{1/3}, i=4,5
     \right\} 
     \nonumber \\
     && \bigcup \;
     \left.
     \left\{ 
     2^{-1/3} 
     \left\{ 
     \frac{m(\mathrm{E}_6) }{[H(\mathrm{E}_6)]^2} \right\}^{1/3}
     \right\} \right) .
     \label{e-defn-eta}
\end{eqnarray}
As} in the case of $\FOm$ {\Eqdot}(\ref{e-FOm-set}), 
a single value of 
{$2\rinj'$ or $2\rinj \rC$}
can correspond to more than one exact-FLRW
model.

\subsection{Measure spaces and probability spaces}
\label{s-rinj-measures}


Let us set $\Sigma$ to be
the smallest $\sigma$-algebra that contains 
all open sets on ${\FOm}$ induced by the usual topology on $\mathbb{R}$.
Then $(\Frinj, \Sigma,\lambda)$,
where $\lambda$ is the Lebesgue measure,
is a measure space over {$2\rinjdimful$, twice} the 
injectivity {radius} of exact-FLRW
models of the Universe. 
Let us normalise this by defining
\begin{equation}
  \widehat{\lambda} = \frac{\lambda}{  
    \postrefereechanges{
      \left(\frac{8 \pi G }{3}\right)^{1/3} \eta }
    -   10a(\tposttopo) {\hGpc}  },
\label{e-normalise-lambda}
\end{equation}
\postrefereechanges{where $\eta$ is given by 
(\ref{e-defn-eta}),
yielding} the probability space $(\Frinj, \Sigma,\widehat{\lambda})$.

Hence, under the obvious \postrefereechanges{probability} space over the size of the 3-spatial
section of FLRW models, 
the probability of a flat FLRW model is unity and the probability
of a curved FLRW model is zero:
\begin{eqnarray}
\widehat{\lambda} \left( 2\rinj \mid_{\Omega = 1}  \right) &=& 1, \nonumber \\
\widehat{\lambda} \left( 2\rinj \mid_{\Omega \not= 1}  \right) &=& 0.
\end{eqnarray}
\postrefereechanges{This follows from (\ref{e-Frinj-set}) and
(\ref{e-normalise-lambda}), since the Lebesgue measure on an interval
on the real line is the length of that interval, and the Lebesgue 
measure on a set of discrete, isolated points is zero.}

{An event that occurs with a probability of one is referred to 
as being {\em almost sure} (a.s.) in the Kolmogorov 
construction of probability spaces. This should not be confused with certainty.} 
For example, {an event of a uniform random} process
that selects a value from the interval $0 \le x \le 1$ on 
the real line results in {a specific} value $x^*$, despite the
fact that {the process a.s.\footnote{\protect{{\em Almost sure} 
and {\em almost surely} are both abbreviated as {\em a.s.}}}\ does not choose that specific value $x^*$.}
Thus, {an a.s.\ outcome is not certain, even though its probability is one.} 
In the case of interest here, 
if the Lebesgue measure is adopted 
{and} normalised, then
a flat model {a.s.\ occurs} and
a non-flat model {a.s.\ does not occur}.

\subsection{How can $\Omega=1$ {occur almost surely}?}

The probability space over $\Omega$, i.e. $(\FOm, 2^{\FOm},\widehat{\mu})$, 
follows from the link between local and global definitions of the 
density parameter, i.e. between curvature and topology, 
{implied by} Hypothesis~\ref{hyp-m-H-dependonM}. 
This {relation} can be described in terms of 
the curvature radius $\rC$, as shown above. {Thus, equality
of the local and global definitions of $\Omega$ discretises it.}
In the flat case, the curvature is zero and $\rC$ is undefined.
The mathematical nature of {flat,} constant curvature 3-manifolds
allows them a continuous range of fundamental domain shapes.
The physical processes that are presumed to select 
a 3-manifold from among those mathematically available
are not constrained to choose fundamental domain size
parameters from a discrete set.
Unless a physical constraint is added, a continuous,
finite range in $2\rinjdimful$, the size parameter adopted here, 
is allowed.

On the other hand, constant-curvature, curved 3-manifolds do not allow $2\rinj$ to
vary. The same rigidity that suggests the discrete measure in 
$(\FOm, 2^{\FOm},\widehat{\mu})$ leads even further, to the Lebesgue measure in
$(\Frinj, \Sigma,\widehat{\lambda})$. Another way of saying this
is that Hypothesis~\ref{hyp-m-H-dependonM} not only links together the curvature
and topology of exact-FLRW models, so that the rigidity of curved
3-manifolds constrains them to a discrete parameter space, but it also
frees up flat 3-manifolds to occupy a continuous parameter space, 
if the parameter chosen is, for example, the injectivity {radius}.
This is 
 {why, given Hypothesis~\ref{hyp-m-H-dependonM},}
the natural choice of measure leads to a probability
space where a flat model {a.s.\ occurs} 
 {($P \equiv \widehat{\lambda} = 1$)}
and a non-flat model {a.s.\ does not occur}
 {($P \equiv \widehat{\lambda} = 0$)}.

\section{Conclusions} \label{s-conclu}

Cosmic topology has been referred to in the cosmological context in
the pre-relativistic era \citep{Schw00} and by \citet{deSitt17},
\citet{Fried23,Fried24}, \citet{Lemaitre31ell} and \citet{Rob35}
during the founding of relativistic cosmology. When the topology of
the FLRW models is taken into account, the popular idea that
$\Omega=1$ FLRW models constitute a class of measure zero among the
full set of exact-FLRW models is no longer self-evident.  This has
been shown here by \postrefereechanges{proposing that the parameters
  determined at a given epoch $\tposttopo$ following topology
  evolution, i.e. as the results of the processes of primordial spatial
  3-manifold evolution (quantum or otherwise), are the global (by
  volume), total (by component) mass-energy $m(M)\vert_{\tposttopo}$
  and the Hubble parameter $H(M)\vert_{\tposttopo}$, i.e. these are functions of the
  3-manifold $M$ selected by those processes, while the density $\Omega$
  is relegated to a derived parameter
  (Hypothesis~\ref{hyp-m-H-dependonM}).}

This hypothesis leads to quite different measure spaces and associated
probability spaces than those in which  
$\Omega=1$ FLRW models constitute a class of measure zero.
When $\Omega$ is used to parametrise the FLRW
models, the hypothesis leads to the discrete measure as
the obvious choice.
In this case, there is no obvious constraint requiring
$\mu(\Omega=1) = 0$.  Moreover, use of the injectivity {radius
$\rinj$} to parametrise FLRW models, rather than $\Omega$, suggests
the Lebesgue measure as the natural measure.  In this case, the
measure of the class of {{\em non-flat}} models is zero.  Since the
measure is normalisable, the probability of a flat FLRW model in the
corresponding probability space is unity. 
Hence, in this case, a flat model 
{occurs a.s.}\ and a non-flat model
{a.s.\ does not occur, 
in the Kolmogorov probability sense of these terms.}
This is the reverse of what
has been thought to be the case when topology is ignored.

How does this happen? For exact-FLRW models with compact spatial sections,
{the density parameter has both a local 
and a global physical meaning. 
\postrefereechanges{The} requirement of equality between the two 
definitions of the density parameter, and
the} rigidity of curved,
constant-curvature {3-manifolds,} reduce the parameter freedom
of the non-flat models. 
In contrast, the lack of rigidity of the flat models allows them 
a continuum of possible sizes.
Hence, the flat models become much more probable. 
This approach suggests a motivation independent
of inflationary scenarios for the recent work finding that a
$\mathrm{T}^3$ FLRW model provides a good fit to the WMAP sky maps
\citep{WMAPSpergel,Aurich07align,Aurich08a,Aurich08b,Aurich09a}.
Moreover, \postrefereechanges{a} physical relation between the
curvature and topology of comoving space \postrefereechanges{is implied}.

\ack
Thank you to Karolina Zawada, Bartosz Lew 
\postrefereechanges{and anonymous referees for several}
useful comments.
Some of this work was carried out within the framework of the European
Associated Laboratory ``Astrophysics Poland-France''.
Use was made 
of 
the computer algebra program {\sc maxima},
the GNU {\sc Octave} command-line, high-level numerical computation software 
(\url{http://www.gnu.org/software/octave}),
and
the
Centre de Donn\'ees astronomiques de Strasbourg 
(\url{http://cdsads.u-strasbg.fr}).








\fhex
\fHW

\postrefereestart
\appendix
\section{Injectivity radii}
\subsection{Injectivity radius of hexagonal-prism spaces}
\label{s-app-hex}
Figure~\ref{f-hex} shows a projection of the fundamental domain (FD) of a
hexagonal-prism space, with dimensions that maximise the injectivity
radius for a fixed volume, i.e. the ``vertical'' and ``horizontal'' 
shortest closed geodesics are both of length $L$. The volume of the
FD is $V= \sqrt{3} L^3/2$, so
\begin{equation} 
\frac{2\rinj}{V^{1/3}} =  2^{1/3} 3^{-1/6}.
\end{equation}

\subsection{Injectivity radius of Hantzsche-Wendt space}
\label{s-app-HW}
Figure~\ref{f-HW} shows a projection of the FD of a
regular (side length $L/2 := L_{\mathrm{a}}/2 = L_{\mathrm{e}}/2 = L_{\mathrm{u}}/2$) 
Hantzsche-Wendt space, $E_6$ in Table~I of \citet{RiazFlat03}. Five of the
square-pyramidal extensions around the inscribed cube, A, B, D, E, and the
unshown lower square-pyramidal extension can be cut off and pasted
around the sixth one C, making a second cube of identical size. Hence,
the FD volume is 
\begin{equation}
V = 2 (L/2)^3.
\end{equation}
Let us, w.l.o.g., choose holonomy $g$ to be the first mapping
in (48) in \protect\citet{RiazFlat03}
\begin{equation}
g : (x,y,z) \mapsto (x + L/2, -y +L/2, -z ).
\label{e-defn-g}
\end{equation}
Since the shift
in the $x$ direction is identical for all points in the FD,
$2\rinj$ cannot be smaller than $L/2$. 
The point $p_1 = (L/4, L/4, 0)$ in Fig.~\ref{f-HW} is mapped
to $g(p_1) = (3L/4,L/4,0)$, i.e. by a distance of $L/2$.
Thus, the $L/2$ lower bound is attained.
Hence, 
\begin{equation} 
\frac{2\rinj}{V^{1/3}} = \frac{L/2} {2^{1/3} (L/2)} = 2^{-1/3}.
\end{equation}

\postrefereestop

\clearpage

\end{document}